\documentclass[prd,aps,twocolumn,a4paper,showkeys,nofootinbib,superscriptaddress]{revtex4-2}

\UseRawInputEncoding
\usepackage{graphicx}
\usepackage[dvipsnames]{xcolor}
\usepackage{dcolumn}
\usepackage{bm}
\usepackage{hyperref}
\usepackage[mathlines]{lineno}
\usepackage{amsmath,amssymb,euscript}
\usepackage{calrsfs}
\allowdisplaybreaks

\usepackage{tikz}
\usetikzlibrary{calc,decorations.markings}
\usetikzlibrary{snakes}
\usepackage{rotating}
\usepackage{comment}
\usepackage{empheq}
\usepackage{ulem}

\newcommand{\CH}{\mathcal{H}}

\newcommand{\spa}{\ , \ \ }

\allowdisplaybreaks[2]
\numberwithin{equation}{section}	
\renewcommand{\theequation}{\arabic{section}.\arabic{equation}}

\definecolor{darkgreen}{rgb}{0.0, 0.55, 0.1}

\begin{document}

\title{Strong-gravity precession resonances for binary systems\\ orbiting a Schwarzschild black hole}

\author{Marta Cocco}
\email{marta.cocco@nbi.ku.dk}
 \affiliation{Dipartimento di Fisica e Geologia, Universit\`a di Perugia, I.N.F.N. Sezione di Perugia, \\ Via Pascoli, I-06123 Perugia, Italy}
 \affiliation{Center of Gravity, Niels Bohr Institute, Copenhagen University,\\ Blegdamsvej 17, DK-2100 Copenhagen \O{}, Denmark}
\author{Gianluca Grignani}
\email{gianluca.grignani@unipg.it}
\affiliation{Dipartimento di Fisica e Geologia, Universit\`a di Perugia, I.N.F.N. Sezione di Perugia, \\ Via Pascoli, I-06123 Perugia, Italy}
\author{Troels Harmark}
\email{harmark@nbi.ku.dk}
 \affiliation{Center of Gravity, Niels Bohr Institute, Copenhagen University,\\ Blegdamsvej 17, DK-2100 Copenhagen \O{}, Denmark}
 \affiliation{Nordita, KTH Royal Institute of Technology and Stockholm University, \\
Hannes Alfv\'ens v\" ag 12, SE-106 91 Stockholm, Sweden
\\ (Authors appear in alphabetical order)} 
\author{Marta Orselli}
\email{marta.orselli@unipg.it}
 \affiliation{Dipartimento di Fisica e Geologia, Universit\`a di Perugia, I.N.F.N. Sezione di Perugia, \\ Via Pascoli, I-06123 Perugia, Italy}
  \affiliation{Center of Gravity, Niels Bohr Institute, Copenhagen University,\\ Blegdamsvej 17, DK-2100 Copenhagen \O{}, Denmark}
\author{Daniele Pica}
\email{daniele.pica@nbi.ku.dk}
 \affiliation{Dipartimento di Fisica e Geologia, Universit\`a di Perugia, I.N.F.N. Sezione di Perugia, \\ Via Pascoli, I-06123 Perugia, Italy}
 %

\begin{abstract}

Binary systems of compact objects in close orbit around a supermassive black hole (SMBH) may form in galactic nuclei, providing a unique environment to probe strong-gravity tidal effects on the binary's dynamics.
 In this work, we investigate precession resonances arising between the periastron precession frequency of a binary system and its orbital frequencies around the SMBH. By modeling the SMBH as a Schwarzschild black hole, we find that relativistic effects in the tidal field give rise to a significantly richer resonance spectrum compared to the Newtonian case. This result is supported by both perturbative and numerical analyses of the quadrupolar tidal interaction in the strong-gravity regime. Our results reveal new signatures for strong-gravity effects in such triple systems, with potential implications for gravitational-wave astronomy.

\end{abstract}

\maketitle
\tableofcontents

\section{Introduction}  
\label{Sec:intro}

Recent detections of gravitational waves (GWs) have revealed a rich population of compact-object binaries that merge on observable timescales~\cite{LIGOScientific:2016dsl, Abbott_2019, LIGOScientific:2020ibl}. 
In the coming decades,
next generation detectors~\cite{Amaro-Seoane:2012aqc, Amaro_Seoane_2023, Punturo:2010zz,berry2019uniquepotentialextrememassratio, amaroseoane2017laserinterferometerspaceantenna, Barack_2019}
will extend the sensitivity into the millihertz regime, enabling the monitoring of binaries over significantly longer timescales. The extended observational reach may reveal binaries that are part of hierarchical triple systems, in which a third body perturbs the binary through tidal forces~\cite{Antonini:2012ad,Samsing:2017rat,Samsing:2017xmd}.

An important question is whether strong-gravity effects in the tidal forces from a black hole perturber are observationally detectable~\cite{zwick2025environmentaleffectsstellarmass}. 
In the strong-gravity regime, treating the black hole as an effective point particle becomes insufficient; instead, a full metric description of the black hole is required.
This possibility could be relevant in future observations, as several mechanisms are known to produce binary systems orbiting close to a supermassive black hole (SMBH).
Examples include the migration trap by accretion disc dynamics~\cite{Bellovary:2015ifg, Secunda:2020mhd, Peng_2021, 2023APS..APRM14001S, Tagawa_2020, zhang2024dynamicsgravitationalwavesignalbinary} and tidal capture processes~\cite{Generozov:2018niv,Chen:2018axp}.
Notably, a binary system of two stellar-mass objects has recently been detected in close orbit around Sgr A*~\cite{Pei_ker_2024}.

In this paper, we explore the phenomenon of {\it precession resonances}~\cite{Kuntz:2021hhm} as a probe of strong-gravity effects in tidal forces. Our setting involves a binary system of compact objects of stellar mass, referred to as the {\it inner binary}, orbiting a SMBH. GW emission from such systems has been studied previously in various contexts~\cite{Chen:2018axp, Chen_2018,Camilloni:2023rra,Camilloni:2023xvf,yin2024relativisticmodelbemrisystems, meng2024gravitationalwavesexcitedbinary, Jiang:2024mdl}. 

The orbit of the inner binary around the SMBH can itself be treated as an effective binary system, referred to as the {\it outer binary}. We assume that the size of the inner binary remains small compared to the Schwarzschild radius of the SMBH.

So far, precession resonances have been identified only in the nearly Newtonian regime, which in our setting corresponds to the outer orbit lying far from the SMBH. 

In this regime, resonances arise from the quadrupole contribution to tidal forces, occurring when the periastron precession frequency of the inner binary (a post-Newtonian (PN) effect) is commensurate with the Newtonian orbital frequency of the outer binary~\cite{Kuntz:2021hhm}.
We write this as
\begin{equation}
    \label{resonanceconditionNewt}
    q\, \dot{\gamma} = p\, \Omega_{\rm{N}} \,,
\end{equation}
with $p$ and $q$ being positive integers, and where $\dot{\gamma}$ is the angular frequency of the inner binary periastron precession and $\Omega_{\rm{N}}$ is the angular frequency of the outer orbit in the Newtonian limit. 

In this paper, we show that strong-gravity effects lead to a richer spectrum of resonances, as multiple frequencies are associated with the outer orbit when the binary system is close to the SMBH.
For simplicity, we model the SMBH as a Schwarzschild black hole.
We show that the resonance condition now becomes
\begin{equation}
    \label{resonanceconditionRel}
    q\, \dot{\gamma} = k\, \Omega_{\hat{r}} + l\,  \Omega_{\hat{\Psi}} \,.
\end{equation}
Here $\Omega_{\hat{r}}$ is the angular frequency associated with the radial motion of the outer orbit and $\Omega_{\hat{\Psi}}$ is the angular frequency associated with the rotation of the local inertial frame of the inner binary, known as Marck's angle~\cite{Marck:1973}. 
The resonance condition in Eq.~\eqref{resonanceconditionRel} involves three integers, with $q$ assumed positive. This leads to a richer spectrum of resonance frequencies, as combinations with negative $k$ are allowed, given that $\Omega_{\hat{r}}\leq \Omega_{\hat{\Psi}}$. 
In the Newtonian limit, both frequencies $\Omega_{\hat{r}}$ and $\Omega_{\hat{\Psi}}$ reduce to $\Omega_{\rm{N}}$, so that $p=k+l$ in Eq.~\eqref{resonanceconditionNewt}.

As the inner binary evolves due to GW emission from radiation-reaction~\cite{1964PhRv..136.1224P}, it will go through these resonances. 

We find that a resonance can cause a significant jump in the eccentricity of the inner orbit, similar to the nearly Newtonian case~\cite{Kuntz:2021hhm}.
The richer spectrum of resonances in our strong-gravity scenario could offer valuable insights into the properties of the surrounding environment~\cite{Hendriks:2024gpp}.

This work complements the study of strong-gravity effects in the von Zeipel-Lidov-Kozai (ZLK) mechanism for the same type of three-body system~\cite{Camilloni:2023xvf, Maeda:2023uyx, Maeda:2023tao}.  
The ZLK mechanism induces high eccentricity of the inner binary through the tidal forces of the perturber~\cite{1910AN....183..345V, Lidov:1962wjn, Kozai:1962zz}, but only if the mutual inclination exceeds approximately $39^{\circ}$. 
The ZLK mechanism operates on timescales that are long compared to the outer orbit period, allowing it to be modeled by averaging over both the inner and outer orbits. In contrast, the precession resonances studied in this paper occur on timescales comparable to the outer orbit period.
In~\cite{Camilloni:2023xvf} strong-gravity effects were shown to significantly alter the evolution of the inner binary, enhancing the efficiency of the ZLK mechanism and accelerating the merger of the two companions.
Additionally, the study demonstrates that properly analyzing a triple system in the strong-gravity regime requires using General Relativity to describe the interaction between the binary and the SMBH.

Among the possible non-secular effects, we focus on resonances, which play a crucial role in GW emission from binary systems interacting with an external environment. Resonances can alter the gravitational waveform emitted by a binary~\cite{PhysRevLett.123.101103, Gupta_2021, Brink_2015, Brink:2015roa, Flanagan_2014, Chandramouli_2022, Gupta_2020, Flanagan:2010cd}, accelerate the merger by enhancing eccentricity~\cite{Nishizawa_2016, Hoang_2018, Naoz_2016, Naoz_2013, Camilloni:2023xvf, Sharpe:2025voe, Bhaskar:2022tzq, Antonini:2012ad}, and even affect the stability of such systems~\cite{Stockinger:2024tai, Stockinger:2025ouz}.

Strong-gravity effects in three-body systems have also been considered in~\cite{Yang:2017aht, Camilloni:2023xvf, Camilloni:2023rra, Grilli:2024fds, Pica:2024ysi}, both for inner binary systems with comparable-mass objects and for extreme-mass-ratio inspirals.

The paper is organized as follows: in Sec.~\ref{sec:Hamiltonian}, we derive the Hamiltonian for the inner binary of compact objects orbiting a SMBH. 
In Sec.~\ref{analytic}, we introduce the relevant frequencies for our analysis and, in the small-eccentricity limit of the outer orbit around the SMBH, perform a Fourier expansion of the inner binary's Hamiltonian---computed at quadrupole order---explicitly in terms of these frequencies.   This yields a novel analytical condition for the precession resonance. In Sec.~\ref{numerical}, we numerically study the evolution of the inner binary, showing that strong-gravity effects lead to a richer spectrum of resonances that can significantly influence its dynamics.  Finally, in Sec.~\ref{sec:conclusion}, we summarize the main results and discuss directions for future work.

\section{Hamiltonian} \label{sec:Hamiltonian}

In this paper, we consider a binary system of compact objects orbiting a SMBH. 
We model the binary system as two point particles with masses $m_1$ and $m_2$, referring to it as the {\it inner binary} and its trajectory as the {\it inner orbit}.
The compact objects in the inner binary may be stellar-mass black holes or neutron stars. In the case where the compact objects are black holes, we require that their separation $r$ is much larger than their Schwarzschild radii, ensuring that the dynamics can be treated approximately using Newtonian mechanics. Moreover, we include the 1PN correction to describe periastron precession.

The SMBH is modeled as a Schwarzschild black hole of mass $M_*$, with metric 
\begin{equation}
\label{schwarzschildmetric}
    ds^2 = - \left( 1 - \frac{2 G M_* }{c^2\hat{r}} \right) c^2 d\hat{t}^2 + \frac{d\hat{r}^2}{1-\frac{2 G M_* }{c^2\hat{r}}  } + \hat{r}^2 d\hat{\Omega}^2\,,
\end{equation}
where $d\hat{\Omega}^2=d\hat{\theta}^2+\sin^2\hat{\theta} d\hat{\phi}^2$.
The orbit of the inner binary around the
Schwarzschild black hole is referred to as the {\it outer orbit}.

We adopt the {\it small-tide} approximation, which assumes that the characteristic size $r$ of the inner binary is much smaller than the curvature radius $\mathcal{R}$ associated with the geometry generated by the SMBH~\cite{Poisson:2009qj}. This ensures that the influence of the SMBH can be modeled through tidal forces. 
Along a geodesic, the requirement $r\ll \mathcal{R}$ implies~\cite{Camilloni:2023xvf}
\begin{equation}
\label{small_binary_condition_general}
r \ll \hat{r} \sqrt{\frac{c^2 \hat{r}}{2 G M_*}}\, ,
\end{equation}
where $\hat{r}$ is the separation between the center of mass of the inner binary and the SMBH.
This condition can be satisfied in two possible regimes. In the first, $\hat{r} \gg 2 G M_*/c^2$, the binary lies far from the SMBH, and its tidal influence can be treated within the Newtonian approximation~\cite{Kuntz:2021hhm, Randall:2018qna}. In the second regime, the binary orbits in close proximity to the SMBH, at distances of a few Schwarzschild radii,
where strong-gravity effects must be taken into account. In this case, Eq.~\eqref{small_binary_condition_general} reduces to
\begin{equation}
\label{small_binary}
r \ll \frac{2 G M_*}{c^2}\, ,
\end{equation}
i.e., the inner binary separation must remain well below the Schwarzschild radius of the SMBH.

\begin{figure*}[ht]
    \centering
\includegraphics[width=1\textwidth]{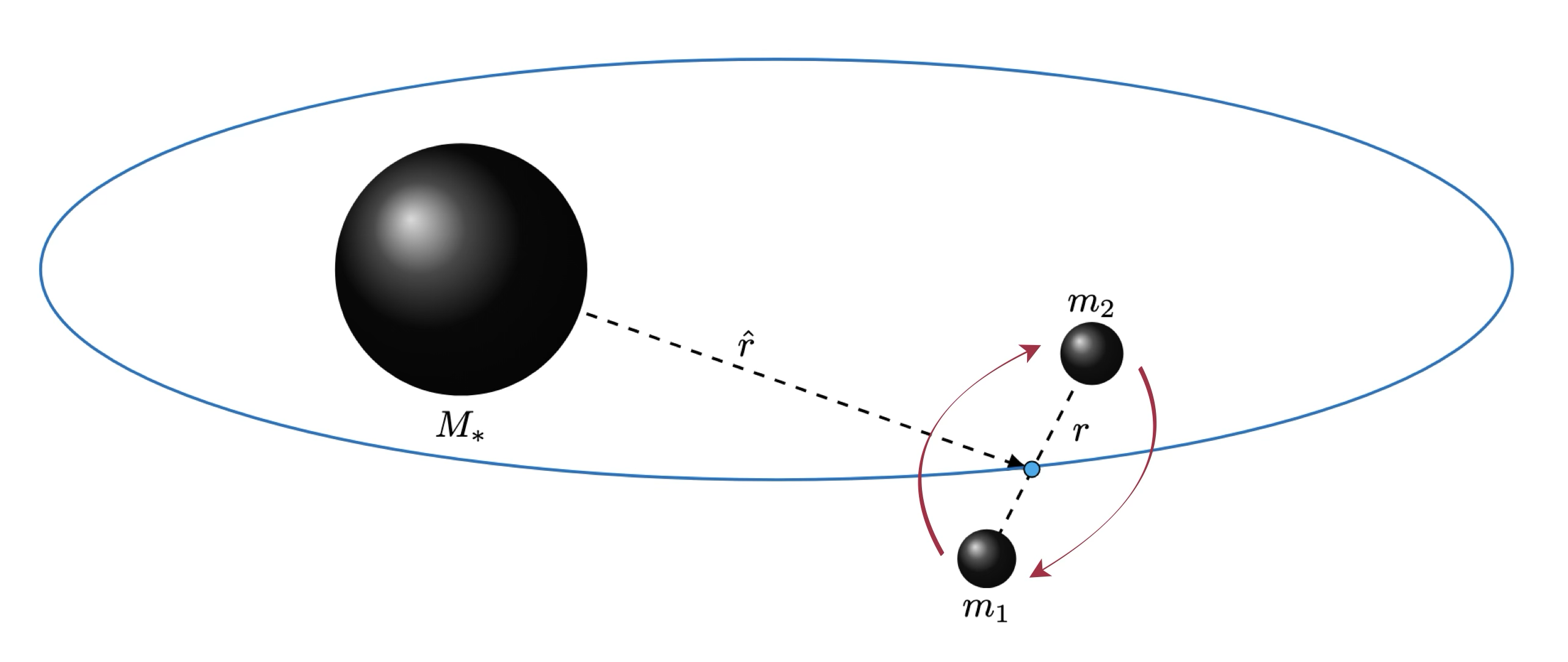}
    \caption{\textit{Schematic representation of the triple system.} 
  Two compact objects with masses $m_1$ and $m_2$ form the inner binary (red orbit) with characteristic size $r$. The center of mass of the inner binary (cyan dot) follows a geodesic around a non-rotating SMBH of mass $M_*$ at a distance $\hat{r}$, defining the outer orbit (blue). 
    }
    \label{fig:configuration}
\end{figure*}

In this work, we focus on the strong-field regime and model the tidal field at quadrupole order. 
Since we allow for outer orbits close to the SMBH, the small-tide condition in Eq.~\eqref{small_binary} is satisfied by enforcing a hierarchy in the masses, namely $M_*\gg m_1, m_2$~\cite{Camilloni:2023xvf}. A schematic representation of the hierarchical triple system is shown in Fig.~\ref{fig:configuration}.

We are interested in the dynamics of the triple system on timescales longer than that of the inner orbit. 
On such timescales, the inner binary can be effectively treated as a finite-size particle orbiting the Schwarzschild black hole, thus reducing the triple system to an effective two-body problem~\cite{Kuntz:2021ohi,Kuntz:2022onu}. We refer to this effective binary system as the {\it outer binary}. 

The Hamiltonian that describes this system is given by
\begin{equation}
\label{total_ham}
\CH = \CH_{\rm pp} + \CH_{\rm inner} + \CH_{\rm quad} \,.
\end{equation}
Here, $\CH_{\rm pp}$ denotes the Hamiltonian of a relativistic point particle, i.e., a structureless particle that does not interact with the curvature of the Schwarzschild spacetime. It is given by
\begin{equation}
\label{Hpp}
\begin{split}
    & \CH_{\rm pp} = \frac{1}{2} g^{\mu \nu} p_\mu p_\nu = -\frac{p_{\hat{t}}^2}{c^2} \left(1-\frac{2 G M_* }{c^2\hat{r}} \right)^{-1}  \\ & + \left(1- \frac{2 G M_* }{c^2\hat{r}} \right) p_{\hat{r}}^2 + \frac{p_{\hat{\theta}}^2}{\hat{r}^2}+\frac{p_{\hat{\phi}}^2}{\hat{r}^2 \sin^2\hat{\theta}}\,.\\
    \end{split}
\end{equation}
This Hamiltonian describes the leading-order geodesic motion of the inner binary's center of mass as it moves along the outer orbit. Without loss of generality, we can restrict the motion to the equatorial plane, so that $\hat{\theta}=\tfrac{\pi}{2}$ and $p_{\hat{\theta}}=0$. 
The two constants of motion associated with the geodesic are denoted $\hat{E}$, representing the energy per unit mass, and $\hat{L}$, representing the angular momentum per unit mass.
The geodesic motion is parametrized by the proper time $\hat{\tau}$ of the outer orbit.

The second term in Eq.~\eqref{total_ham}, $\CH_{\rm inner}$, describes the internal dynamics of the inner binary. Since we are interested in timescales longer than the period of the inner orbit, this term is averaged over the inner orbit. Consequently, this part of the Hamiltonian can be interpreted as describing the internal structure of the effective finite-size particle.
As we focus on quadrupole interactions, the binary system can be treated as freely falling. This means that, to leading order,  its center of mass moves on a geodesic of the background spacetime~\cite{Camilloni:2023xvf, Maeda:2023tao}.

Under these assumptions, one finds~\cite{Camilloni:2023xvf,Kuntz:2021ohi} 
\begin{equation}
\label{Hinner}
    \CH_{\rm inner} = - \frac{GM\mu}{2a} - 3\mu \frac{G^2 M^2}{a^2 c^2 \sqrt{1-e^2}}\,.
\end{equation} 
We briefly review the quantities appearing in this expression. 
The first term in Eq.~\eqref{Hinner} corresponds to the Newtonian binding energy of the binary, while the second term describes the general relativistic periastron precession, averaged over the inner orbit~\cite{Kuntz:2021ohi}. This precession arises as a 1PN correction and occurs on a timescale $t_{\rm 1PN}\sim 1/\dot{\gamma}$.
In Eq.~\eqref{Hinner}, $M$ and $\mu$ denote, respectively, the {\it total mass} and the {\it reduced mass} of the inner binary, defined as
\begin{equation}
    M = m_1+m_2 \spa \mu = \frac{m_1m_2}{M} \,.
\end{equation}
Moreover, the inner orbit is parametrized by the {\it semi-major axis} $a$ and {\it eccentricity }$e$ using the relation
\begin{equation}
\label{rparametr}
    r = \frac{{a}(1-{e}^2)}{1+{e}\cos {\psi}} \,,
\end{equation}
where $\psi$ is the {\it true anomaly}, the angle between the direction of the periastron and the current orbital position.
The orientation of the inner binary relative to the outer orbit is specified by the Euler angles $(\gamma,I,\vartheta)$, where $\gamma$ is the {\it argument of periapsis}, $I$ the {\it orbital inclination}, and $\vartheta$ the {\it longitude of the ascending node}.
The binary's motion can be conveniently described using the action-angle {\it Delaunay variables}. The angle variables are
$(\beta,\gamma,\vartheta)$, with $\beta$ the {\it mean anomaly}, and their conjugate action variables are
\begin{equation}
J_\beta = \mu \sqrt{GMa}~,~~ J_\gamma= J_\beta \sqrt{1-e^2}~,~~J_\vartheta = J_\gamma \cos I~.
\end{equation}

The averaging of Eq.~\eqref{Hinner} over the inner orbit is performed by integrating over one period of the mean anomaly $\beta$, which is the angle that increases uniformly in time. This average is more conveniently expressed as an integral over the true anomaly $\psi$, using the relations between $\beta$, $\psi$, and the {\it eccentric anomaly} $\zeta$
\begin{equation}
\label{keplerslawsinner}
    \cos\psi= \frac{\cos \zeta - e}{1-e \cos \zeta}\,, \quad \beta= \zeta-e \sin\zeta\,.
\end{equation} 
The second expression in Eq.~\eqref{keplerslawsinner} corresponds to Kepler's equation. These relations hold unambiguously because the inner binary follows Newtonian dynamics.
Consequently, for a generic quantity $\mathcal{A}$, its average over the inner orbit is given by~\cite{Camilloni:2023xvf}
    \begin{equation}
    \label{inner-average}
       \left< \mathcal{A} \right> = \frac{1}{2\pi} \int_0^{2\pi} \mathcal{A}\, d \beta = \frac{1}{2\pi} \int_0^{2\pi} \mathcal{A} \, \frac{(1-e^2)^{3/2}}{(1+e \cos \psi )^2} \, d\psi\,.
    \end{equation} 

Finally, the third term in Eq.~\eqref{total_ham}, $\mathcal{H}_{\rm quad}$, describes the coupling of the inner binary to the quadrupole tidal field of the Schwarzschild background. Like the previous term, it is averaged over the inner orbit and can thus be interpreted as a coupling between the curvature of the background spacetime and the internal structure of the effective finite-size particle. The unaveraged form of this Hamiltonian was derived in~\cite{Camilloni:2023xvf} for the more general case of a Kerr black hole; in this work, we focus on the special case of a Schwarzschild black hole. Performing the inner-orbit average as defined in Eq.~\eqref{inner-average} yields
\begin{equation}
\begin{split}
    \label{Hquadinneraveraged}
    \CH_{\rm quad} &= \frac{G M_*}{\hat{r}^3} \frac{\mu a^2}{2} \left[ \frac{2+ 3 e^2}{2} \right. \\
    &+ 3 \frac{\hat{L}^2}{c^2 \hat{r}^2} \frac{2+3 e^2 - 5e^2 \cos 2\gamma}{4} \sin^2 I \\
    &- 3 \left(1+\frac{\hat{L}^2}{c^2 \hat{r}^2}\right) \left( \frac{2+3 e^2 + 5e^2 \cos 2\gamma}{4} \cos^2(\hat{\Psi}-\vartheta) \right. \\
    &+ \frac{2+3 e^2 - 5e^2 \cos 2\gamma}{4} \sin^2(\hat{\Psi}-\vartheta)\cos^2 I  \\ 
    & \left. \left. + \frac{5 e^2}{2} \sin 2 \gamma  \cos(\hat{\Psi}-\vartheta)\sin(\hat{\Psi}-\vartheta)\cos I \right) \right] \,.
    \end{split}
\end{equation}
Here, $\hat{\Psi}$ denotes {\it Marck's angle}, which parametrizes the rotation of the inertial frame that is parallel-transported along the geodesic. In Schwarzschild spacetime, it obeys~\cite{Marck:1973}
\begin{equation}
\label{Psidot}
   \frac{d\hat{\Psi}}{d\hat{\tau}}= \frac{\hat{E}\hat{L}}{c^2 \hat{r}^2+\hat{L}^2} \,.
\end{equation}
Eq.~\eqref{Hquadinneraveraged}
is expressed in the  inertial reference frame of Marck~\cite{Marck:1973} as outlined in~\cite{Camilloni:2023xvf}. In Sec.~\ref{distantstar}, we discuss its form in an alternative frame, the {\it distant-star frame}.

To connect to Ref.~\cite{Kuntz:2021hhm}, we note that one can also express $\CH_{\rm quad}$ as
\begin{equation}
    \label{Hquad}
    \CH_{\rm quad}= \frac{1}{2} c^2 {Q}^{ij} \mathcal{E}_{ij}~,~~~{\rm with}~~(i,j)=1,2,3~.
    \end{equation}
    Here $Q^{ij}$ is the traceless quadrupole moment of the inner binary, averaged over the inner orbit, and is given by~\cite{Kuntz:2021hhm}
\begin{equation}
\label{qij}
  Q^{ij}= \frac{\mu a^2}{2}\left[(1+4e^2) \, u^i u^j + (1-e^2)\, v^i v^j -\frac{2+3e^2}{3} \,\delta^{ij}\right] \, 
\end{equation}
with 
\begin{equation}\hat{\bm{u}}=R_z(\vartheta) R_x(I) R_z(\gamma)\, \hat{\bm{x}}~,~~\hat{\bm{v}} =R_z(\vartheta) R_x(I) R_z(\gamma) \,\hat{\bm{y}}~,
\end{equation}
where $R_x$ and $R_z$ are rotation matrices, and we have introduced the unit vectors $\hat{\bm{x}}= \left(1,0,0\right)$,  $\hat{\bm{y}}= \left(0,1,0\right)$ and $\hat{\bm{z}}= \left(0,0,1\right)$.

The electric tidal moments $\mathcal{E}_{ij}$ in \eqref{Hquad} encode the full general relativistic coupling to the curvature of the Schwarzschild spacetime~\cite{Marck:1973, Camilloni:2023xvf}. For $\hat{\theta}=\pi/2$, the nonvanishing components are 
\begin{equation}
\label{electrictidal}
    \begin{split}
    \mathcal{E}_{11}=& \left[1-3 \left( 1 + \frac{\hat{L}^2}{c^2 \hat{r}^2}\right) \cos^2 \hat{\Psi}\right] \frac{G M_*}{c^2 \hat{r}^3} \,,\\
    \mathcal{E}_{22}=& \left[1-3 \left( 1 + \frac{\hat{L}^2}{c^2 \hat{r}^2}\right) \sin^2 \hat{\Psi}\right] \frac{G M_*}{c^2 \hat{r}^3} \,,\\
    \mathcal{E}_{33}=& \left( 1 + 3 \frac{\hat{L}^2}{c^2 \hat{r}^2} \right) \frac{G M_*}{c^2 \hat{r}^3} \,,\\
    \mathcal{E}_{12}=& -3 \left( 1 + \frac{\hat{L}^2}{c^2 \hat{r}^2}\right) \frac{G M_*}{c^2 \hat{r}^3} \cos\hat{\Psi} \sin\hat{\Psi} \,.
     \end{split}
\end{equation}
%

\section{Perturbative analysis}  
\label{analytic}

In this section, we derive the resonance condition given in Eq.~\eqref{resonanceconditionRel}, which describes resonances between the periastron precession frequency of the inner binary and the fundamental frequencies of the outer orbit. 
This derivation, performed in the limit of small outer-orbit eccentricity, $\hat{e} \ll 1$, reveals a richer spectrum of resonances arising from strong-gravity effects. In Sec.~\ref {numerical}, we show numerically that \eqref{resonanceconditionRel} remains valid also for finite $\hat{e}$.


\subsection{Fundamental frequencies of the outer orbit}

The geodesic motion of the outer orbit is governed by the Hamiltonian given in Eq.~\eqref{Hpp}. Since the motion is bounded, it is characterized by three fundamental frequencies $\omega_{\hat{t}}$, $\omega_{\hat{r}}$, and $\omega_{\hat{\phi}}$, associated with the coordinates $\hat{t}$, $\hat{r}$, and $\hat{\phi}$, of the Schwarzschild metric in Eq.~\eqref{schwarzschildmetric}. The fourth coordinate $\hat{\theta}$ does not introduce an independent frequency, as the system's spherical symmetry allows one to restrict the motion to the equatorial plane, $\hat{\theta}=\pi/2$.

It is also important to note that the Hamiltonian in Eq.~\eqref{total_ham} does not depend on the azimuthal coordinate $\hat{\phi}$. Instead, it depends explicitly on Marck's angle $\hat{\Psi}$, which characterizes the rotation of the parallel-transported inertial reference frame along the geodesic. This angle introduces an additional fundamental frequency, $\omega_{\hat{\Psi}}$, which is independent of the three orbital frequencies $\omega_{\hat{t}}$, $\omega_{\hat{r}}$, and $\omega_{\hat{\phi}}$, associated with the geodesic motion~\cite{vandeMeent:2019cam}.

The frequency $\omega_{\hat{t}}$ accounts for gravitational time dilation between the proper time $\hat{\tau}$, which is the local time of the inner binary, and $\hat{t}$, the time measured by an asymptotic observer. Since we wish to express the resonance condition from the perspective of an asymptotic observer, we translate the fundamental frequencies to their asymptotically measured values, as observed at spatial infinity
\begin{equation}
    \Omega_{\hat{r}} = \frac{\omega_{\hat{r}}}{\omega_{\hat{t}}} \spa
    \Omega_{\hat{\phi}} = \frac{\omega_{\hat{\phi}}}{\omega_{\hat{t}}} \spa
    \Omega_{\hat{\Psi}} = \frac{\omega_{\hat{\Psi}}}{\omega_{\hat{t}}} \,.
\end{equation}

Employing the action-angle variable formalism, one can find angle variables such that 
\begin{equation}
    \label{genangleeom}
  q_{\mu}=\Omega_{\mu} \hat{t}\, ~,
\end{equation}
with $\mu=\hat{r}, \hat{\phi}, \hat{\Psi}$.

The geodesic motion can now be solved by expressing $\hat{r}$, $\hat{\phi}$, and $\hat{\Psi}$, in terms of the angles introduced in Eq.~\eqref{genangleeom}. This procedure has been developed for the more general Kerr spacetime for example in~\cite{Hinderer:2008dm, Schmidt:2002qk, Bini:2016iym, vandeMeent:2019cam}, and the Schwarzschild case is recovered by setting the black hole spin parameter to zero.

Note that only two fundamental frequencies, $\Omega_{\hat{r}}$ and $\Omega_{\hat{\Psi}}$, enter the resonance condition. This follows directly from the fact that---as it will be confirmed below---only the generalized coordinates $q_{\hat{r}}$ and $q_{\hat{\Psi}}$ appear in the Hamiltonian~\eqref{total_ham}. In a generalization to a Kerr black hole, an additional fundamental frequency would arise, associated with the polar motion $\hat{\theta}$.

In the following, we consider the small-eccentricity limit of the outer orbit, $\hat{e} \ll 1$, and  expand $\hat{r}$ in powers of $\hat{e}$, treating it as a function of the generalized angles $q_{\mu}$:
\begin{equation}
\label{hatr_exp}
\hat{r}=  \hat{a} \left(1 - \hat{e} \cos q_{\hat{r}}\right) +\mathcal{O}(\hat{e}^2) \,.
\end{equation}
For Marck's angle, the expansion reads
\begin{equation}
    \label{hatPsi_exp}
        \hat{\Psi} =  q_{\hat{\Psi}} + 2 \hat{e} \, \frac{\hat{\sigma}-4}{\hat{\sigma}-2} \sqrt{\frac{\hat{\sigma}-3}{\hat{\sigma}-6}} \sin q_{\hat{r}} +\mathcal{O}(\hat{e}^2) \,
\end{equation}
where, for convenience, we introduced the dimensionless quantity 
\begin{equation}
    \hat{\sigma} = \hat{a} \frac{c^2}{G M_*}
    \,. 
\end{equation}

For small eccentricity $\hat{e}$, we also have~\cite{Hinderer:2008dm, Schmidt:2002qk,Bini:2016iym, vandeMeent:2019cam}
\begin{subequations}
    \begin{align}
    \label{omegarsmalle}
    \Omega_{\hat{r}}&= \Omega_{\rm N} \sqrt{\frac{\hat{\sigma}-6}{\hat{\sigma}}}+\mathcal{O}(\hat{e}^2) \,,
    \\
    \label{omegaPsismalle}
    \Omega_{\hat{\Psi}} & = \Omega_{\rm N} \sqrt{\frac{\hat{\sigma}-3}{\hat{\sigma}}} +\mathcal{O}(\hat{e}^2) \,.
    \end{align}
\end{subequations}
Here, $\Omega_{\rm N} = \sqrt{G M_* / \hat{a}^3}$  denotes the Newtonian fundamental frequency associated with the Keplerian motion of the outer orbit. In the Newtonian limit, where $\hat{\sigma} \to \infty$, we have $\Omega_{\hat{r}}, \Omega_{\hat{\Psi}} \to \Omega_{\rm N}$. For finite $\hat{\sigma}$ and bound orbits with $\hat{e} < 1$, we have numerically verified that $\Omega_{\hat{r}} < \Omega_{\hat{\Psi}}$.

\subsection{Precession resonances}
\label{sec:precession}

Considering the Hamiltonian in Eq.~\eqref{total_ham}, it is evident that resonances between the 1PN periastron precession and the fundamental frequencies of the outer orbit arise solely from the $\CH_{\rm quad}$ term. This term encodes the interaction between the inner binary and the quadrupole tidal field generated by the Schwarzschild black hole. 
In fact, neither $\CH_{\rm pp}$ nor $\CH_{\rm inner}$ exhibit explicit dependence on angles or angle variables. For $\CH_{\rm pp}$, this is because the angle variables \eqref{genangleeom} solve the geodesic equations. For $\CH_{\rm inner}$, which governs the 1PN periastron precession, the absence of angle dependence results from averaging over the inner orbit.

To identify the resonances arising from $\CH_{\rm quad}$, we expand \eqref{Hquadinneraveraged}  in the small outer eccentricity limit, $\hat{e}\ll 1$, substituting $\hat{r}$ and $\hat{\Psi}$ using Eqs.~\eqref{hatr_exp} and \eqref{hatPsi_exp}, respectively.
Keeping only terms that involve trigonometric functions of $2 \gamma - (k\, \Omega_{\hat{r}} + l\Omega_{\hat{\Psi}}) \,\hat{t}$, which are the terms that contribute to the resonances, or terms that are constant, we obtain
an expansion of the form~\footnote{Terms not included in this expansion oscillate rapidly and thus average out over time~\cite{Kuntz:2021hhm}.} 
\begin{equation}
   \label{HREL2}
   \begin{split}
       & \CH_{\rm quad}=\frac{G a^2 M_* \mu}{48 \hat{a}^3}\Big[ (2+3e^2)\, h_{0,0}(I,\vartheta)  \\[2mm]
       & + 15 e^2 \sum_{k,l\neq (0,0)} \Big( f_{k,l}(I,\vartheta) \cos\xi_{k,l} + g_{k,l}(I,\vartheta) \sin \xi_{k,l}\Big) \Big] \,,
   \end{split}
\end{equation}
with resonance angles
\begin{equation}
    \label{resonant_angles}
    \xi_{k,l}\equiv 2 \gamma - (k\, \Omega_{\hat{r}} + l\, \Omega_{\hat{\Psi}})\,\hat{t}
\end{equation}
and Fourier coefficients of the form
\begin{subequations}
\label{coefficientsHtildeRel}
    \begin{align}
        f_{k,l}(I,\vartheta)=&  \Big( a_{k,l}\, (1+ \cos^2 I) + 2 {b}_{k,l} \cos I \Big) \cos 2\vartheta \nonumber\\
        &
        + c_{k,l} \sin^2 I \,,
        \\
        g_{k,l}(I,\vartheta)=& -\Big({b}_{k,l} \, (1+ \cos^2 I)+ 2a_{k,l} \cos I  \Big) \sin 2\vartheta \,,
    \end{align}
\end{subequations} 
while $h_{0,0}$ is
\begin{equation}
        h_{0,0}(I,\vartheta)= \frac{3\hat{\sigma}}{3-\hat{\sigma}}(3 \cos^2 I - 1) + \mathcal{O}\left(\hat{e}^2\right)\,.
\end{equation}
This last term does not contribute to the resonances, but it is important to accurately model the evolution of the inner binary.
Note that $l$ is restricted to the values
\begin{equation}
    l = -2,0,2 \,.
\end{equation}
This restriction arises from the fact that $\hat{\Psi}$ appears in the electric tidal moments \eqref{electrictidal} only through $\cos (2\hat\Psi)$ or $\sin (2\hat\Psi)$.
We also assume that $\dot{\gamma}=d\gamma/d\hat{t}$, $\Omega_{\hat{r}}$ and $\Omega_{\hat \Psi}$ are positive. As a result, resonances can only occur for values of $k$ and $l$ such that $k\, \Omega_{\hat{r}} + l\, \Omega_{\hat{\Psi}}>0$. 

At first order in the expansion for small $\hat{e}$, the non-zero Fourier 
coefficients are
\begin{equation}
    \label{fouriercoefficientsk1l0}  
    c_{1,0} = -\frac{9}{2} \frac{\hat{\sigma}+2 }{\hat{\sigma}-3} \, \hat{e} + \mathcal{O}\left(\hat{e}^2\right) \,,
\end{equation}
\begin{equation}
     \label{fouriercoefficientskm1l2}  
     \begin{split}
    & a_{-1,2}={b}_{-1,2}=  \\
    & = \left[\frac{12-9 \hat\sigma}{4 (\hat{\sigma}-3)}  +  \frac{3 (\hat\sigma - 4)}{\sqrt{(\hat\sigma - 6)(\hat\sigma - 3)}} \right] \hat{e}+ \mathcal{O}\left(\hat{e}^2\right) \,, 
   \end{split}
\end{equation}
\begin{equation}
\label{fouriercoefficientsk0l2}
    a_{0,2}= {b}_{0,2} =  -\frac{3}{2}\frac{\hat{\sigma} - 2 }{\hat{\sigma}-3} + \mathcal{O}\left(\hat{e}^2\right) \,,
\end{equation}
\begin{equation}
     \label{fouriercoefficientsk1l2}  
     \begin{split}
    & a_{1,2}={b}_{1,2}=  \\
    & = \left[\frac{12-9 \hat\sigma}{4 (\hat{\sigma}-3)}  -  \frac{3 (\hat\sigma - 4)}{\sqrt{(\hat\sigma - 6)(\hat\sigma - 3)}} \right] \hat{e}+ \mathcal{O}\left(\hat{e}^2\right) \,. \end{split}
\end{equation}
This confirms the presence of resonances of the form
$q\, \dot{\gamma} = k\, \Omega_{\hat{r}} + l\,  \Omega_{\hat{\Psi}} $
 between the periastron precession frequency of the inner binary and the fundamental frequencies of the outer orbit, as anticipated in Eq.~\eqref{resonanceconditionRel} in the introduction. 
 The resonances occur with $q=2$ reflecting the structure of the quadrupole moment $Q^{ij}$ in Eq.~\eqref{qij} of the inner binary. The coupling to the tidal field $\mathcal{E}_{ij}$, described by \eqref{Hquad} and \eqref{electrictidal}, determines the allowed values of $(k,l)$. The above Fourier coefficients, at first order in $\hat{e}$, show that the following resonances occur
\begin{subequations}
    \begin{align}
    \label{res10}
    2 \dot{\gamma} & =  \Omega_{\hat{r}} \,,
    \\
    \label{resm12}
    2\dot{\gamma} &= - \Omega_{\hat{r}}+2\,\Omega_{\hat{\Psi}} \,,
    \\
    \label{res20}
    2 \dot{\gamma} & =  2\,\Omega_{\hat{\Psi}} \,,
    \\
    \label{res12}
    2\dot{\gamma} &=  \Omega_{\hat{r}}+2\,\Omega_{\hat{\Psi}}\,.
    \end{align}
\end{subequations}
Non-zero Fourier coefficients with $k + l \geq 4$ arise only at second and higher orders in $\hat{e}$, leading to additional resonances with $k + l \geq 4$.
Resonances with $k + l \leq 0$ are excluded, as they do not satisfy the condition $k\, \Omega_{\hat{r}} + l\, \Omega_{\hat{\Psi}} > 0$.

 Comparing with the Newtonian precession resonance condition \eqref{resonanceconditionNewt}, we observe that in the Newtonian limit ($\hat \sigma \rightarrow \infty$) 
 one has that $p = k + l$.
 In this limit in fact, both $\Omega_{\hat{r}}$ and $\Omega_{\hat{\Psi}}$ tend toward the Newtonian frequency $\Omega_{\rm N}$.
 Therefore, the inclusion of  strong-gravity effects in the tidal interaction between the inner binary and the Schwarzschild black hole leads to the splitting of certain Newtonian resonances into multiple distinct relativistic resonances. This splitting is evident for $p=1$, which separates into $(k,l)=(1,0)$ and $(k,l)=(-1,2)$, corresponding to Eqs.~\eqref{res10} and \eqref{resm12}, respectively. 
Similarly, the resonances given in Eqs.~\eqref{res20} and \eqref{res12} correspond to $p = 2$ and $p = 3$ in the Newtonian limit, respectively.

Note that our results are valid for finite $\hat{\sigma}$, corresponding to the strong-gravity regime near the Schwarzschild black hole. This approach goes beyond the inclusion of perturbative PN corrections, as it incorporates the full relativistic structure of the tidal coupling.
In this regime, the resonances given in Eqs.~\eqref{res10}--\eqref{res12} are fully distinct and well separated.

\section{Numerical analysis of resonances} 
\label{numerical}

In this section, we complement the perturbative analysis of precession resonances from the previous section with a comprehensive numerical study.
As we will show, this numerical approach confirms that the resonance condition given in Eq.~\eqref{resonanceconditionRel} remains valid even for finite eccentricity $\hat{e}$ of the outer orbit.

We numerically solve the evolution equations for the orbital parameters of both the inner and outer orbits, without resorting to a Fourier expansion of the Hamiltonian in Eq.~\eqref{Hquadinneraveraged}. 
In this way, we can track the evolution of the inner binary in the strongly relativistic regime---specifically, for orbits where the binary minimum distance to the SMBH is only a few Schwarzschild radii---while also accounting for finite eccentricity $\hat e$
of the outer orbit.

To numerically solve the dynamics of the outer binary, we use the outer orbit variables introduced in Sec.~\ref{sec:Hamiltonian}, rather than the action-angle formalism used in Sec.~\ref{analytic}.
To describe outer orbit geodesics with finite eccentricity $\hat e$, we
follow~\cite{Chandrasekhar:1985kt} in parametrizing the radial coordinate $\hat{r}$ in terms of the {\it relativistic anomaly} $\hat \psi$ as
\begin{equation}
\label{rhatparametr}
    \hat{r} = \frac{\hat{a}(1-\hat{e}^2)}{1+\hat{e}\cos \hat{\psi}} \,,
\end{equation}
where $\hat e$ plays the role of the eccentricity and $\hat a$ that of the semi-major axis of the outer orbit, in analogy with Newtonian dynamics. With this parametrization, one obtains the following evolution equation for the relativistic anomaly~\cite{Chandrasekhar:1985kt}
\begin{equation}
    \label{DpsiDt}
    \begin{split}
    \frac{d \hat{\psi}}{d \hat{t}}= & \sqrt{\frac{G M_*}{\hat{a}^{3}(1-\hat{e}^2)^{3}}} \frac{ \left(1+\hat{e} \cos \hat{\psi}\right)^{2}}{\sqrt{(2 \delta -1)^2- 4 \delta^2 \hat{e}^2}} \\ 
    & \times \sqrt{1-2 \delta (3+ \hat{e} \cos \hat{\psi})} \left(1-2\delta (1+\hat{e} \cos\hat{\psi})\right) \,,
    \end{split}
\end{equation}
where $\delta= G M_* / (c^2 \hat{a} (1-\hat{e}^2) )$. This equation provides the evolution of $\hat r$ through Eq.~\eqref{rhatparametr}.
Note that we evolve the triple system using the asymptotic time $\hat t$, since this is the natural time for an observer far away from the SMBH.

Considering the quadrupole tidal force contribution to the Hamiltonian \eqref{Hquadinneraveraged}, the only additional outer-orbit quantity for which we require an evolution equation is Marck's angle $\hat{\Psi}$. This angle evolves according to 
\begin{equation}
    \label{DPsiDt}
     \frac{d\hat{\Psi}}{d\hat{t}}= \frac{1}{u^{\hat{t}}}\frac{\hat{E}\hat{L}}{c^2 \hat{r}^2+\hat{L}^2} \,,
\end{equation}
which is the redshifted version of Eq.~\eqref{Psidot}, with $u^{\hat{t}}=d\hat{t} / d\hat{\tau}$ being the redshift factor. This completes the description of the outer orbit.

Turning to the evolution of the inner binary physical quantities, it is governed by the Lagrange Planetary Equations \eqref{lagrangeequations} as presented in App.~\ref{AppA}, which arise from the terms $\CH_{\rm inner}+\CH_{\rm quad}$ in the Hamiltonian \eqref{total_ham}. However, for this dynamics, we also include radiation-reaction effects, as they provide a natural mechanism for the system to evolve through the various resonances. Accordingly, we add the following contributions to the Lagrange Planetary Equations~\eqref{lagrangeequations} for the inner semi-major axis $a$ and eccentricity $e$
\begin{equation}
\label{radiationreaction}
\begin{split}
    \left(\frac{da}{d\hat{t}}\right)_{\rm RR}=& -\frac{1}{u^{\hat{t}}}\frac{64}{5} \frac{G^3 \mu M^2}{c^5 a^3} \frac{\left( 1+ \frac{73}{24}e^2 + \frac{37}{96} e^4\right)}{\left(1-e^2\right)^{7/2}}  \,, \\
    \left(\frac{de}{d\hat{t}}\right)_{\rm RR}=& -\frac{1}{u^{\hat{t}}}\frac{304}{15} \frac{G^3 \mu M^2}{c^5 a^4} \frac{e\left(1+\frac{121}{304}e^2\right)}{\left(1-e^2\right)^{5/2}} \,.
\end{split}   
\end{equation}
These contributions describe the loss of energy and angular momentum in the averaged evolution of an isolated inner binary~\cite{PhysRev.131.435,1964PhRv..136.1224P}. They can be consistently added since radiation-reaction effects enter as a 2.5PN correction.

This completes the set of evolution equations for the triple system. 
In summary, Eqs.~\eqref{rhatparametr}, \eqref{DpsiDt}, and \eqref{DPsiDt} describe the outer-orbit dynamics, including the rotation of the inertial reference frame.
The Lagrange Planetary Equations \eqref{lagrangeequations},  with the radiation-reaction terms \eqref{radiationreaction} added, govern the evolution of the inner orbit. 
Together, these form a system of coupled ordinary differential equations in the asymptotic time $\hat t$, which can be solved numerically for given initial values of the parameters $a,~e,~I,~\gamma,~\vartheta,~\hat{\Psi},~\hat{\psi}$ and fixed values of $\hat{e},~ \hat{a}$.~\footnote{We neglect GW emission from the outer orbit, allowing us to treat $\hat a$ and $\hat e$ as constants during the evolution of the inner binary.}

We recall that our focus is on the evolution of the inner binary in the strong-gravity regime, which occurs when the inner binary is close to the SMBH.
We consistently enforce the small-tide approximation introduced in Eq.~\eqref{small_binary}, ensuring that the inner binary remains sufficiently tight compared to the curvature radius of the outer spacetime. Furthermore, as we aim to describe astrophysically relevant scenarios, we require the triple system to remain stable when the inner binary is placed close to the SMBH. Specifically, we always set the semi-major axis $\hat a$ of the outer orbit larger than the semi-major axis $\hat a_{\rm ISO}$ of the Innermost Stable Orbit (ISO), i.e.,~$\hat a > \hat a_{\rm ISO}$, which for a non-spinning black hole of mass $M_*$ is given by~\cite{Brink:2015roa}

\begin{equation}
    \label{ISO}
    \hat{a}_{\rm ISO}=  \left(\frac{6+2 \hat{e}}{1-\hat{e}^2} \right) \frac{G M_*}{c^2} \,.
\end{equation}
Additionally, we ensure that the inner binary system is stable against tidal disruption events caused by the SMBH. This requires imposing the condition~\cite{Miller_2005,Antonini:2012ad}
\begin{equation}
    \label{ritide}
    (1-\hat{e})\hat a \gtrsim 
    a \left(\frac{3 M_*}{M}\right)^{1/3}.
\end{equation}
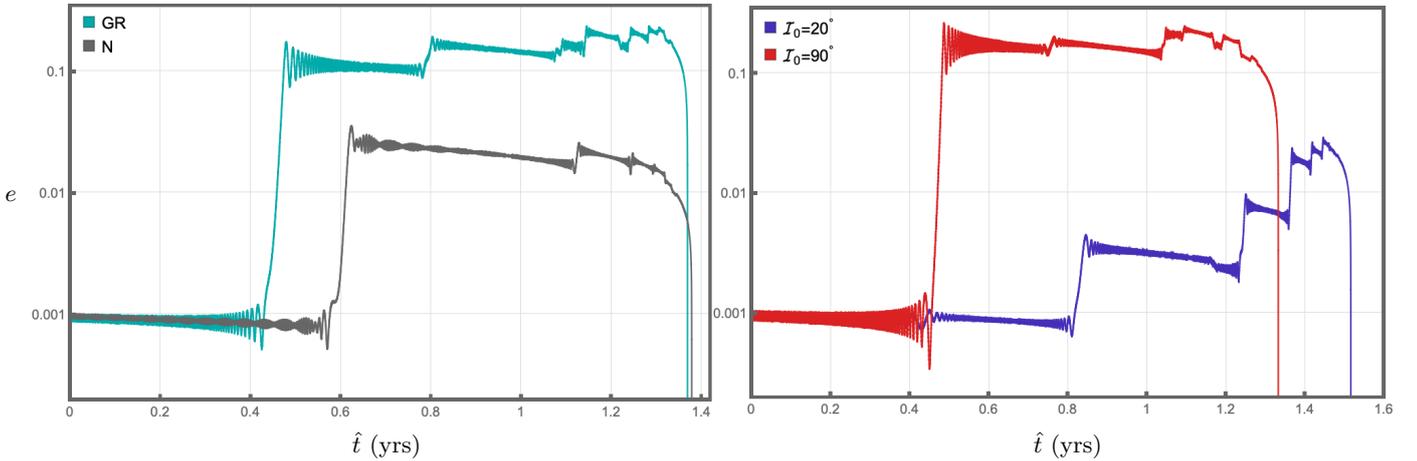
\begin{figure*}[ht]
    \centering
    \includegraphics[width=1\textwidth]{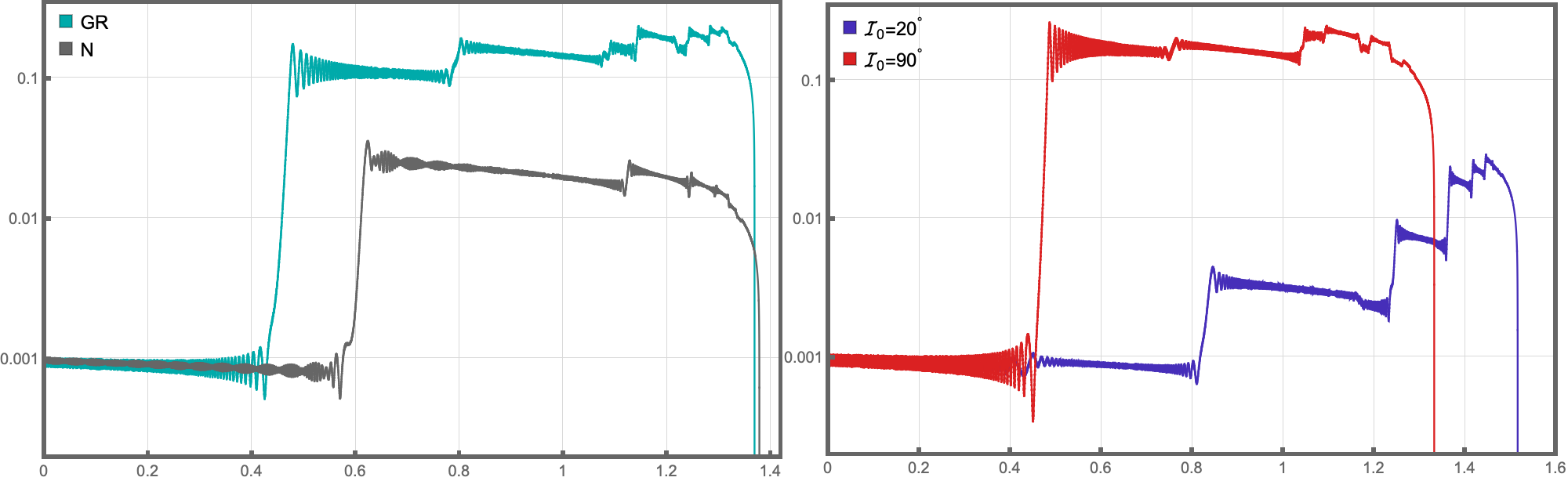}
    \begin{picture}(0,0)
        \put(-135,-2){\small $\hat t$ (yrs) }
        \put(120,-2){\small $\hat t$ (yrs) }
        \put(-265,93){$e$}
    \end{picture}
    \caption{Left panel: Comparison 
    between the numerical results obtained with our approach in the regime of strong gravity (GR) and the standard Newtonian description (N) for an initial inclination $I_0=55^{\circ}$.
    Right panel: Numerical results in the regime of strong gravity for different values of the initial inclination. 
    In both panels, we used
    the following parameters and initial conditions for the inner binary: total mass $M = 50~M_{\odot}$, reduced mass $\mu=12.5 ~M_{\odot}$, semi-major axis $a_0 \sim 0.0006~\rm AU$, inner eccentricity $e_0 = 0.001$ and $\gamma_0= \vartheta_0 = 0^{\circ}$. 
    For the outer binary we used: $M_* = 4 \times 10^6~M_{\odot}$, semi-major axis $\hat{a} = 18~G M_* / c^2 \sim 0.7~\rm AU$ and outer eccentricity $\hat{e}=0.4$. 
    With these parameters, in the left panel the inner binary merges in $ \sim 1.3$ yrs. In the right panel, the merger time is about $1.37$ yrs for the red curve and $\sim 1.5$ yrs for the blue curve. 
    Moreover, the peak frequency associated with the inner binary's emission of GWs lies in the LISA bandwidth for the entire lifetime of the inner binary, where we considered $0.001~\text{Hz} < f_{\rm GW}^{\rm LISA} < 0.1~\text{Hz}$. Finally, the ratio between the ZLK and inner precession timescales is $t_{\rm ZLK}/t_{\rm 1PN}\sim 40$.}
    \label{fig1_compIxx}
\end{figure*}

In Fig.~\ref{fig1_compIxx}, we present numerical evolutions of the triple system for two different sets of initial conditions. Specifically, the left panel compares our fully relativistic approach (cyan curve) with an approximately Newtonian treatment of the triple system (gray curve), where the SMBH is also treated as a point particle. This comparison clearly shows that, in the strong-gravity regime, the inner binary encounters more resonances, resulting in more jumps in its eccentricity. As discussed in Sec.~\ref{analytic}, this behavior arises from the additional fundamental frequencies describing the outer orbit in the strong-gravity regime.
Furthermore, the left panel shows that strong gravitational effects significantly amplify 
 the eccentricity $e$ of the inner binary: in our relativistic description, $e$ increases by roughly two orders of magnitude from its initial value $e_0$, whereas in the Newtonian case, the increase is only about an order of magnitude.

It is also worth noting that the peak frequency of the emitted GWs, given approximately by~\cite{Wen_2003}
\begin{equation}
    \label{eq:peakfrequency}
    f_{\rm GW}\approx \frac{\sqrt{GM}}{\pi \left[a(1-e^2)\right]^{3/2}}(1+e)^{1.1954} \,,
\end{equation}
remains within the LISA band $(0.001 \rm{Hz} - 0.1 \rm{Hz})$ throughout the entire evolution of the inner binary system depicted in Fig.~\ref{fig1_compIxx}, both in the relativistic and Newtonian cases. This has a direct impact on the gravitational waveform, causing a shift in its phase at each resonance encountered by the inner binary during its evolution~\cite{PhysRevLett.123.101103, Gupta_2021, Flanagan_2014, Chandramouli_2022, Gupta_2020, Hendriks:2024gpp, Lynch:2024ohd, Samsing:2024syt, Flanagan:2010cd}. Therefore, to avoid biases when modeling the effects of resonances on the gravitational waveform of a binary system orbiting close to a SMBH, it is crucial to adopt a description of the triple system that fully incorporates strong gravitational effects. This renders the Newtonian approximation insufficient for accurate predictions. 
Finally, the occurrence of multiple resonance peaks within the LISA band provides a compelling opportunity to discriminate precession resonance effects from other potential resonances, as highlighted in~\cite{Kuntz:2021hhm}. 

In the right panel of Fig.~\ref{fig1_compIxx}, we illustrate how the initial mutual inclination between the inner and outer orbits affects the amplitude of the resonances. Specifically, an inner binary with a high initial inclination (red curve) undergoes a substantially greater enhancement in eccentricity upon encountering resonances. This, in turn, results in a shorter merger time compared to an otherwise identical binary with a lower initial inclination (blue curve). It is important to emphasize that throughout our analysis we enforce the condition $t_{\rm ZLK} \gg t_{\rm 1PN}$, thus suppressing the eccentricity oscillations induced by the ZLK mechanism, even in cases of high initial inclinations, where such effects would typically be expected to arise.

Thus far, we have used the procedure presented in this section to study the evolution of the inner binary in the case of a finite outer eccentricity $\hat e$. However, we can also
perform a numerical study for small values of $\hat e$, allowing for a direct comparison between the numerical results and the analytical predictions derived in Sec.~\ref{analytic}. 
The results of this comparison are displayed in Fig.~\ref{NumAnalyicalComp}. The curves represent solutions to the Lagrange Planetary Equations provided in Appendix~\ref{AppA}, supplemented by the radiation-reaction equations given in Eq.~\eqref{radiationreaction}. The blue curve corresponds to the outcome of the numerical analysis for small values of $\hat{e}$, while the pink curve is obtained by solving the same equations using the Fourier-expanded Hamiltonian in Eq.~\eqref{HREL2}. This Hamiltonian, along with the Fourier coefficients in Eqs.~\eqref{fouriercoefficientsk1l0}--\eqref{fouriercoefficientsk1l2} and the orbital frequencies in Eqs.~\eqref{omegarsmalle} and~\eqref{omegaPsismalle}, is obtained through a systematic expansion in small outer eccentricity, $\hat{e} \ll 1$.

 For completeness, we also include the results of a numerical integration performed within the Newtonian approximation, shown as the gray curve in Fig.~\ref{NumAnalyicalComp}. From this comparison, we observe that the two distinct resonance peaks in the strong-gravity regime, associated with $(k,l)=(1,0)$ and $(k,l)=(-1,2)$ in Eq.~\eqref{resonanceconditionRel}, merge into a single peak in the Newtonian case, corresponding to $p=1$ in Eq.~\eqref{resonanceconditionNewt}.

Fig.~\ref{NumAnalyicalComp} illustrates that the analytical model developed in Sec.~\ref{analytic} provides an accurate description of the dynamical evolution of the inner binary in the regime of small outer eccentricity $\hat{e}$. More specifically, the model successfully predicts the location of each resonance encountered by the inner binary during its evolution, thereby confirming the validity of the underlying resonance condition. 
Furthermore, it accurately reproduces the amplitude of these resonances, confirming the accuracy of the Fourier coefficients given in Eqs.~\eqref{fouriercoefficientsk1l0}--\eqref{fouriercoefficientsk1l2}. 
The discrepancy between the analytical and numerical results near the resonance at $(k,l)=(0,2)$ is presumably due to the growing significance of higher-order corrections in $\hat{e}$,
particularly terms of order
$\mathcal{O}(\hat{e}^2)$.

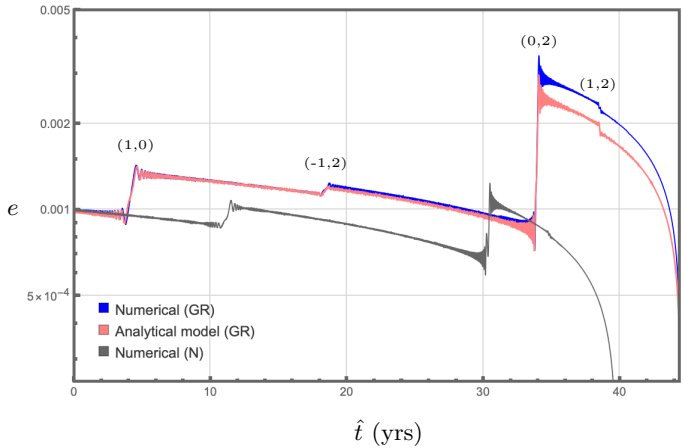
\begin{figure}[h]
    \centering
\includegraphics[width=0.48\textwidth]{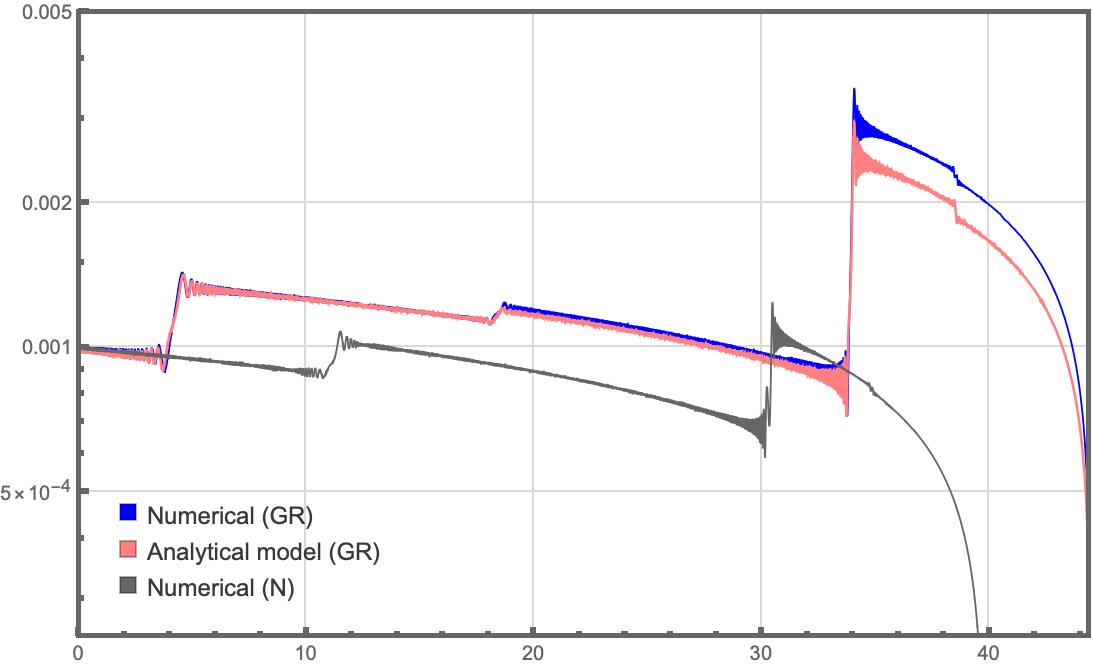}
    \begin{picture}(0,0)
        \put(0,-5){\small $\hat t$ (yrs) }
        \put(-89,105){\tiny (1,0)}
        \put(-19,98){\tiny (-1,2)}
        \put(62,144){\tiny (0,2)}
        \put(84,128){\tiny (1,2)}
    \end{picture}
    \caption{ 
    Comparison between the analytical model developed in Sec.~\ref{analytic} and the numerical results obtained in the strong-gravity regime for small outer eccentricity ($\hat{e} \ll 1$).  The parameters and initial conditions for the inner binary are as follows:  
    total mass $M = 50~M_{\odot}$, 
    reduced mass $\mu=12.5 ~M_{\odot}$, semi-major axis $a_0 \sim 0.0014~\rm AU$, eccentricity $e_0 = 0.001$, initial inclination $I_0 = 60^\circ$ and $\gamma_0= \vartheta_0 = 0^{\circ}$. The inner binary orbits a SMBH of mass $M_* = 5 \times 10^7~M_{\odot}$, with semi-major axis $\hat{a} = 15~G M_* / c^2 \sim 7~\rm AU$ and outer eccentricity $\hat{e}=0.05$.
    As shown in the plot, there is a very good agreement between the analytical prediction (pink curve) and the full numerical integration presented in this section (blue curve). In contrast, the numerical evolution obtained using a Newtonian point-particle approximation (gray curve) exhibits clear deviations, highlighting the importance of incorporating strong-gravity effects in modeling such systems.
    }
    \label{NumAnalyicalComp}
\end{figure}
%

\subsection{Distant-star frame}
\label{distantstar}

The Hamiltonian \eqref{total_ham} is formulated in Marck's local inertial reference frame, which provides a natural basis for describing the dynamics of the inner binary and its coupling to the tidal field of the Schwarzschild black hole.

Alternatively, to capture the global point of view of an asymptotic observer, who can directly measure the Fokker-de Sitter gyroscope precession induced by the outer orbit of the inner binary, one may adopt the {\it distant-star} frame of reference \cite{Bini:2016iym,Camilloni:2023xvf}. This
frame is periodic with respect to the global angle $\hat{\phi}$ of the Schwarzschild metric \eqref{schwarzschildmetric} rather than Marck's angle $\hat{\Psi}$. This corresponds to rotating Marck's frame with an angle given by $\hat{\phi}-\hat{\Psi}$.
As a result of this frame rotation, the longitude of ascending nodes in the distant-star frame transforms as $\vartheta^*=\vartheta+\hat{\phi}-\hat{\Psi}$, while the angles $(\beta, \gamma)$ remain the same.
This change of frame yields a new Hamiltonian $\CH^*=\CH_{\rm pp}+ \CH^*_{\rm inner}+ \CH^*_{\rm quad}$ with
\begin{equation}
\label{Hinner_ds}
    \CH_{\rm inner}^* = - \frac{GM\mu}{2a} - 3\mu \frac{G^2 M^2}{a^2 c^2 \sqrt{1-e^2}}
    + \Omega_{\rm g} J_{\vartheta} \,,
\end{equation} 
where the final term arises from a second-type canonical transformation, and $\Omega_{\rm g} \equiv d (\hat{\phi} - \hat{\Psi})/d\hat{\tau}$ is the gyroscope precession frequency~\cite{Camilloni:2023xvf}.
The quadrupole interaction term in the distant-star frame, $\CH^*_{\rm quad}$, can be obtained directly from $\CH_{\rm quad}$  in Eq.~\eqref{Hquadinneraveraged} by replacing everywhere $\hat{\Psi}-\vartheta$ with $\hat{\phi}-\vartheta^*$. Thus, $\CH^*_{\rm quad}=\CH_{\rm quad}$ since $\hat{\Psi}-\vartheta=\hat{\phi}-\vartheta^*$.

The last term in Eq.~\eqref{Hinner_ds} implies that $\vartheta^*$ evolves in time even in the absence of the tidal interaction encoded in $\CH^*_{\rm quad}$.  This intrinsic time dependence must be accounted for when performing the Fourier expansion of $\CH^*_{\rm quad}$.  Accordingly, although the analysis is formulated in the distant-star frame, the expansion naturally involves the frequency associated with $\hat{\Psi}$, as in Eqs.~\eqref{HREL2} and \eqref{resonant_angles}, rather than the global Schwarzschild frequency tied to $\hat\phi$. As a result, the perturbative analysis in the distant-star frame reproduces the same resonance spectrum as that obtained in Marck's frame, as presented in Sec.~\ref{analytic}.

For the numerical analysis, one may directly employ the full Hamiltonian in the distant-star frame to derive the corresponding evolution equations. We have verified numerically that this approach yields the same set of resonances as those identified in Sec.~\ref{numerical}, thereby confirming the consistency of the two frames.

\section{Conclusions}
\label{sec:conclusion}

In this paper, we have shown that strong-gravity effects significantly modify the precession resonances that can occur in hierarchical triple systems consisting of a compact binary orbiting a SMBH.
By modeling the SMBH as a Schwarzschild black hole, we accounted for the additional fundamental frequencies associated with the relativistic outer orbit. These lead to a richer and more intricate resonance spectrum, as characterized by Eq.~\eqref{resonanceconditionRel}, compared to the approximately Newtonian case, Eq.~\eqref{resonanceconditionNewt}, considered in Ref.~\cite{Kuntz:2021hhm}, where the perturbing SMBH was modeled as a point mass.

Our analysis is valid within the regime defined by the small-tide approximation \eqref{small_binary}, and further constrained by the requirement that the inner binary remains stable against tidal disruption, as enforced by the condition \eqref{ritide}.

An alternative approach to analyzing a hierarchical triple system is to model the inner binary as a spinning particle with fixed multipole moments, following Refs.~\cite{Ruangsri:2015cvg, Zelenka:2019nyp} and effectively describe the system as an EMRI. To include precession resonances in this framework, one must account for the tidal interaction between the primary and secondary companions in the EMRI system through the evolution of the multipole moments, an effect that has not yet been included in previous analyses.

A natural and compelling extension of this work would be to generalize the analysis by modeling the central SMBH as a Kerr black hole. In this scenario, the resonance condition would be modified to include the additional fundamental frequency associated with motion in the polar angle $\hat \theta$, thereby further enriching the spectrum of allowable resonances. 
Such a generalization would not only yield a more complete characterization of the resonant structure but also enable a systematic investigation of the influence of the spin of the SMBH on the secular evolution and resonant dynamics of the inner binary.

Another promising direction for future work would be to incorporate the emission of GWs originating from the outer binary's motion.
This extension could be implemented using the formalism developed in~\cite{yin2024relativisticmodelbemrisystems}, which provides a relativistic framework for modeling GW emission in extreme-mass-ratio inspiral systems.

Moreover, in this work, we developed an analytical framework for characterizing precession resonances by performing a Fourier expansion of the quadrupolar Hamiltonian of the inner binary in the limit of small outer eccentricity $\hat{e}$, retaining terms up to linear order. A natural extension of this analysis would be to systematically include higher-order terms in $\hat{e}$, thereby enabling a more accurate description of the inner binary's dynamics and revealing additional features of the resonance structure.

Finally, a promising direction for future investigation involves extending the present analysis to incorporate the contributions of magnetic quadrupole tidal moments \cite{Poisson:2009qj, Camilloni:2023rra} in the tidal Hamiltonian. In a strong-gravity regime, these contributions may become non-negligible and could significantly affect the evolution of the inner binary, potentially giving rise to additional resonant phenomena.


\section*{Acknowledgments}

We thank J. Samsing for stimulating and interesting discussions. We thank the anonymous referee for relevant feedback. We are indebted to A. Trani and L. Zwick for invaluable comments and discussions. We also thank A. Kuntz for useful feedback.   
M. Cocco, G. Grignani, M. Orselli, and D. Pica acknowledge financial support from the Italian Ministry of University and Research (MUR) through the program ``Dipartimenti di Eccellenza 2018-2022" (Grant SUPER-C). G. Grignani and M. Orselli acknowledge support from ``Fondo di Ricerca d'Ateneo"  2023 (GraMB) of the University of Perugia. G. Grignani, M. Orselli, and D. Pica acknowledge support from the Italian Ministry of University and Research (MUR) via the PRIN 2022ZHYFA2, GRavitational wavEform models for coalescing compAct binaries with eccenTricity (GREAT).
M. Cocco, T. Harmark and M. Orselli acknowledge support by the ``Center of Gravity", which is a Center of Excellence funded by the Danish National Research Foundation under grant No. 184.
T. Harmark thanks University of Perugia for its kind hospitality.



\appendix 
\section{Lagrange Planetary Equations}
\label{AppA}
In this appendix, we write down the Lagrange Planetary Equations used in Sec.~\ref{numerical} for the numerical evaluation.
\renewcommand{\theequation}{\thesection.\arabic{equation}}
\begin{subequations}
\label{lagrangeequations}
\begin{align}
    \frac{d a}{d \hat{t}}=& - \frac{1}{u^{\hat{t}}}\sqrt{\frac{4 a }{G M}} \frac{\partial \tilde{\mathcal{H}}}{\partial \beta} \,, \\
    \frac{d e}{d \hat{t}}=& \frac{1}{u^{\hat{t}}}\left(\sqrt{\frac{1-e^2}{G M a e^2}} \frac{\partial \tilde{\mathcal{H}}}{\partial \gamma} - \frac{1-e^2}{\sqrt{G M a}e} \frac{\partial \tilde{\mathcal{H}}}{\partial \beta}\right)\,, \\
    \frac{d I}{d \hat{t}}=& \frac{1}{u^{\hat{t}}} \frac{1}{\sqrt{G M a (1-e^2)}\sin I}\Bigg( \frac{\partial \tilde{\mathcal{H}}}{\partial \vartheta} - \cos I \frac{\partial \tilde{\mathcal{H}}}{\partial \gamma}\Bigg) \,, \\
    \frac{d \beta}{d \hat{t}}=& \frac{1}{u^{\hat{t}}} \left(\sqrt{\frac{4 a }{G M}} \frac{\partial \tilde{\mathcal{H}}}{\partial a} + \frac{1-e^2}{\sqrt{G M a}e} \frac{\partial \tilde{\mathcal{H}}}{\partial e} \right)\,, \\
    \frac{d \gamma}{d \hat{t}}=&  \frac{1}{u^{\hat{t}}} \left(-\sqrt{\frac{1-e^2}{G M a e^2}} \frac{\partial \tilde{\mathcal{H}}}{\partial e} + \frac{\cot I}{\sqrt{G M a (1-e^2)}}\frac{\partial \tilde{\mathcal{H}}}{\partial I}\right) \,, \\
    \frac{d \vartheta}{d \hat{t}}=& - \frac{1}{u^{\hat{t}}}\frac{1}{\sqrt{G M a (1-e^2)}\sin I}\frac{\partial \tilde{\mathcal{H}}}{\partial I}\,,  
    \end{align}
\end{subequations}
where $\tilde{\mathcal{H}}= (\mathcal{H}_{\rm inner}+ \mathcal{H}_{\rm quad})/\mu$, and $u^{\hat{t}}=d\hat{t} / d\hat{\tau}$ is the redshift factor.


\newpage
\bibliographystyle{apsrev4-1}
\bibliography{References}

\end{document}